\documentclass[]{spie}

\usepackage{amsmath,amsfonts,amssymb}
\usepackage{graphicx}
\usepackage[colorlinks=true, allcolors=blue]{hyperref}

\title{The Automated Data Extraction, Processing, and Tracking System for CHARIS}

\author[a]{Taylor L. Tobin}
\author[a]{Jeffery Chilcote}
\author[b]{Timothy Brandt}
\author[c,d]{Thayne Currie}
\author[e]{Tyler Groff}
\author[d]{Julien Lozi}
\author[d,f,g,h]{Olivier Guyon}

\affil[a]{Department of Physics, University of Notre Dame, Nieuwland Science Hall, Notre Dame, IN, USA}
\affil[b]{Physics Department, University of California Santa Barbara, Broida Hall, University of California, Santa Barbara, CA, USA}
\affil[c]{NASA-Ames Research Center, Moffett Blvd., Moffett Field, CA, USA}
\affil[d]{Subaru Telescope, National Astronomical Observatory of Japan, 650 North A’ohōkū Place, Hilo, HI, USA}
\affil[e]{NASA-Goddard Space Flight Center, Greenbelt, MD, USA}
\affil[f]{Steward Observatory, University of Arizona, Tucson, AZ, USA}
\affil[g]{College of Optical Sciences, University of Arizona, Tucson, AZ, USA}
\affil[h]{Astrobiology Center of NINS, 2-21-1, Osawa, Mitaka, Tokyo, Japan}

\authorinfo{Further author information: (Send correspondence to T.L.T.)\\T.L.T.: E-mail: ttobin2@nd.edu }

\begin{document} 
\maketitle

\begin{abstract}
CHARIS is an IFS designed for imaging and spectroscopy of disks and sub-stellar companions. To improve ease of use and efficiency of science production, we present progress on a fully-automated backend for CHARIS. This Automated Data Extraction, Processing, and Tracking System (ADEPTS) will log data files from CHARIS in a searchable database and perform all calibration and data extraction, yielding science-grade data cubes. The extracted data will also be run through a preset array of post-processing routines. With significant parallelization of data processing, ADEPTS will dramatically reduce the time between data acquisition and the availability of science-grade data products.
\end{abstract}

\keywords{CHARIS, SCExAO, Subaru, direct imaging, software}

\section{INTRODUCTION}
\label{sec:intro} 

The Coronagraphic High Angular Resolution Imaging Spectrograph (CHARIS)\cite{charis5,charis6,charisdrp} is an high-contrast Integral Field Spectrograph (IFS) designed for imaging and spectroscopy of disks, exoplanets, and other sub-stellar companions \cite{Goebel2018,Currie2018,Asensio2019,Currie2019,Currie2020,Lawson2020,Uyama2020}. Located on the Subaru Telescope, it records starlight corrected by the observatory's facility adaptive optics system AO188\cite{ao188} and then further sharpened by the Subaru Coronagraphic Extreme Adaptive Optics system (SCExAO) \cite{scexao}. The spectrograph itself is lenslet-based, supporting observations in $J$, $H$, and $K$ at $\mathcal{R}$ $\sim$ 70 and in a Broadband filter ($\mathcal{R}$ $\sim$ 18) covering $JHK$ simultaneously. It also offers a low-throughput ND filter to avoid saturation when imaging bright sources in lieu of faint companions.

Currently, CHARIS data is sorted manually before undergoing calibration and extraction through the CHARIS Data Reduction Pipeline\cite{charisdrp}. The resulting data cubes can then be manually subjected to PSF subtraction and other post-processing routines, such as KLIP or ALOCI\cite{klip-soummer12,klip-pueyo15,Currie2012}.

To improve ease of use and efficiency of science production, we present progress on the Automated Data Extraction, Processing, and Tracking System (ADEPTS), a fully-automated backend for the CHARIS data system. The backbone of ADEPTS is its processing system, which sorts and logs new CHARIS files as soon as they are detected before running them through the CHARIS DRP and selected post-processing modules for calibration and data extraction.  In addition to its automated data processing system,  ADEPTS includes an improved user interface: all data and reduction information is logged in a searchable database, and all data products created by the automated system are sorted by data type, as well as date, filter, and source. 

In this paper, we discuss the structure of the ADEPT System and progress toward its implementation. In Section \ref{sec:current-system}, we review the current CHARIS data system and process for obtaining science data from raw images. In Section \ref{sec:interface}, we discuss the two aspects of ADEPTS that comprise the improved user interface: a new database for logging and sorting data, as well as tracking data products (Section \ref{ss:database}) and the extended file system for science-quality data products created by the ADEPT System (Section \ref{ss:file-system}). In Section \ref{sec:automation}, we outline the Automated Data Extraction and Processing System that forms the backbone of ADEPTS, and provide a brief update on the current state of the system, as well as an outlook on the system's benefits for future science with CHARIS.

\section{THE CURRENT CHARIS SYSTEM}
\label{sec:current-system}  

There are three types of calibration images to be taken while observing with CHARIS. First, flats of the appropriate filter and spectral resolution are required to be taken the same night as their respective observations. Utilizing the uniform illumination of a tunable halogen flat-field lamp, these flats are used by the CHARIS DRP to calculate nightly shifts in the lenslet PSF, which is required for an accurate wavelength solution\cite{charisdrp}. Broadband and K-band observations also require nightly dark images to correct for the thermal background that appears at K-band wavelengths\cite{charisdrp}; however, this thermal background is negligible in the  J and H bands. Skyflats are also recommended for Broadband and K-band images, to remove the thermal sky background, although their application must be performed manually. If taken, any skyflat images are calibrated and their data extracted as if they were targeted observations; then, the skyflat image cubes are averaged together to create a mean skyflat cube which is manually subtracted from all observational data. For more information on CHARIS's design and performance, see Ref.~\citenum{charis1,charis2,charis3,charis4,charis5,charis6,charis7}.

Under CHARIS's current operational system, all data from CHARIS are stored as FITS files in the system's Network-Attached Storage (NAS). All information on each image, including image type, filter, and target, are stored in the FITS header, while the files themselves are sorted into subdirectories by date. Any data that are marked as bad during observation must be noted manually outside of the FITS header keywords. 

Most CHARIS users, however, do not have direct access to the NAS; they must instead request observation files from someone who does have access. This requires the requester to specify either the individual file numbers as recorded in their observational log or the targets and bands, if multiple observing programs were run on a given night. In the case of the latter, whoever is fulfilling the request must scan the headers of the FITS files taken on the desired date to locate the requested observations, in addition to manually locating any corresponding flat, dark, and skyflat files.  Open use observers usually download CHARIS data from the Subaru Telescope STARS2 archive, using a Python script that retrieves data via \texttt{wget} commands.   CHARIS file sizes can be extremely large, making manual downloading of a full run's set of CHARIS data tedious for both open use observers and those more formally associated with SCExAO/CHARIS teams.

  \begin{figure} [ht]
   \begin{center}
   \begin{tabular}{c}
   \includegraphics[width=0.9\linewidth]{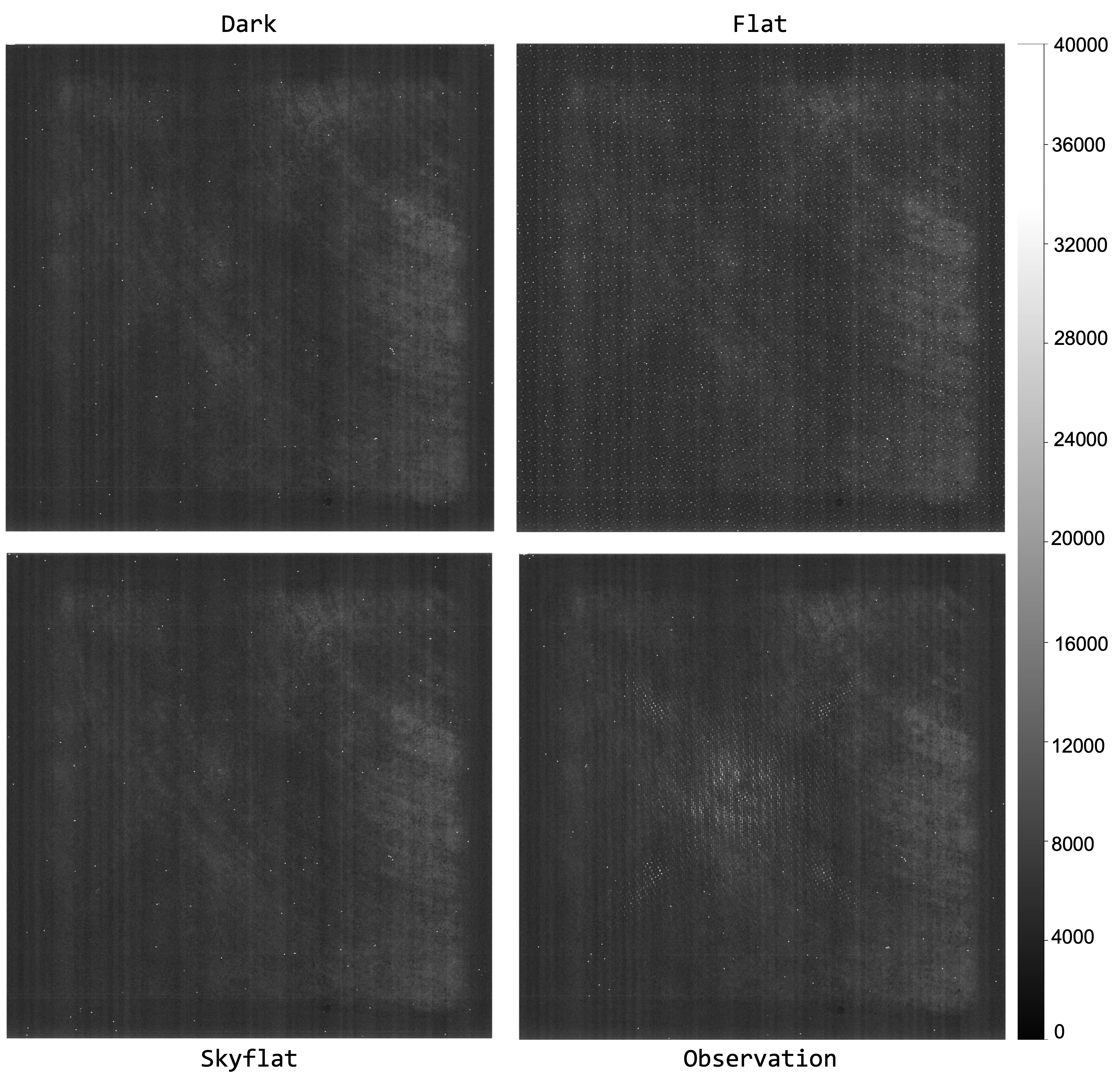}
   \end{tabular}
   \end{center}
   \caption{ \label{fig:sampleims} Samples of the four different types of raw data images from CHARIS: darks, flats, skyflats, and observations, shown in raw counts (see Section \ref{sec:current-system}). In the uniformly-illuminated flat, a grid of lenslets can be seen across the whole image. In the observation, the lenslets trace the PSF. Sample shown are taken with Broadband specifications. }
   \end{figure} 
   
Once the relevant files have been obtained, the data must be calibrated and extracted. This is typically done with the CHARIS DRP, which consists of two modules: \texttt{buildcal}, which processes the dark and flat images into the required calibration files, and \texttt{extractcube}, which uses the built calibration files to extract data cubes from skyflat or observation images. If skyflat images were taken in the appropriate band, the mean skyflat data cube must be subtracted manually from the observations as described above.   Typical run-time for building calibrations is in the tens of minutes range, depending on the number of cores used and processing power.   CHARIS data cubes likewise usually take tens of seconds to several minutes each to extract.  Thus, sequentially converting a full run's worth of CHARIS data to data cubes is also time-intensive.

\begin{figure} [ht]
   \begin{center}
   \begin{tabular}{c}
   \includegraphics[width=0.95\linewidth]{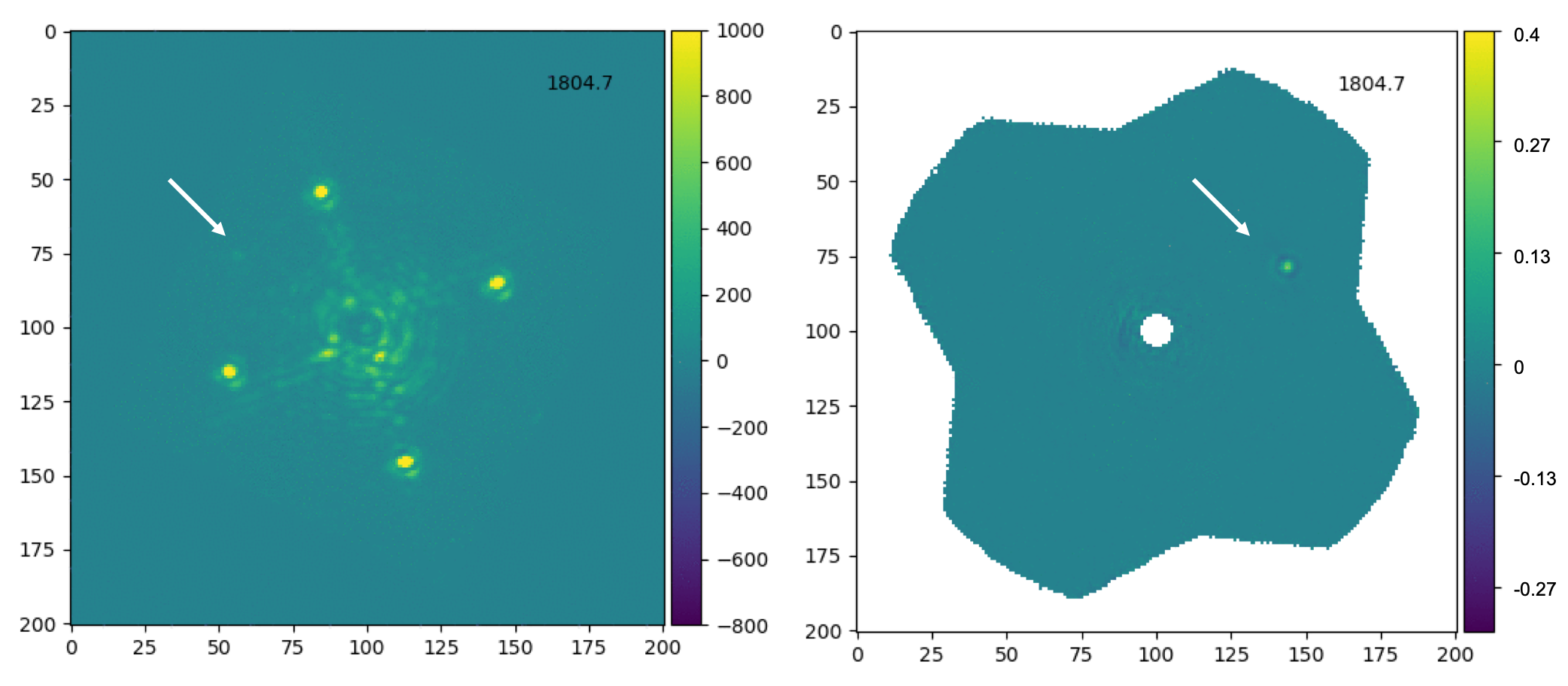}
   \end{tabular}
   \end{center}
   \caption{ \label{fig:sampleext} Example of the extracted data cube of a single Broadband CHARIS observation at one wavelength in units of count rate (left) and the same source after combination with other observations and post-processing with PyKLIP\cite{pyklip} in calibrated contrast units (right). The companion in both images is indicated with a white arrow. Wavelength of the slice is 1804.7 nm in both images. }
   \end{figure}

Finally, it is common practice in direct imaging of disks and substellar companions to follow data extraction with additional image processing. These post-processing methods, including Angular Differential Imaging (ADI)\cite{ADI1,ADI2}, Spectral Differential Imaging (SDI)\cite{SDI1,SDI2},  Locally Optimized Combination of Images (LOCI)\cite{LOCI}, and Karhunen-L\`{o}eve Image Projection (KLIP)\cite{klip-soummer12,klip-pueyo15,pyklip}, seek to remove the stellar PSF, suppress speckle noise, and improve contrast for imaging faint sources and structures (eg. Fig. \ref{fig:sampleext}). These packages are distributed separately and are not affiliated with the CHARIS instrument, so it is up to the user to perform any desired post-processing. Notably, many of these methods require fine-tuning of parameters for optimal PSF subtraction and speckle suppression for a given source.

\section{THE NEW ADEPTS USER INTERFACE}
\label{sec:interface} 

The modifications made to the CHARIS infrastructure as part of the ADEPTS package will seek to improve user experience with a new interface. This consists of a multi-table database for file searching and tracking as well as a sorted file system for all science-grade data products created by ADEPTS.

\subsection{The New CHARIS Database} 
\label{ss:database}  

The CHARIS database, written in SQLite\cite{sqlite}, will allow CHARIS’s users to search for data and browse file information without scanning FITS file headers. There are five primary tables in the database: \textit{CHARISData}, \textit{darks}, \textit{buildcal}, \textit{skyflats}, and \textit{obs}.

The first table, \textit{CHARISData}, houses all information from the FITS headers for every file produced by CHARIS, regardless of image type. In addition, it includes a column for flagging bad data. As discussed further in Section \ref{ss:databaseentry}, this assumes that the observers have created a text file logging the names of all image files that have been manually flagged as bad data. Any files marked as bad will not be included in the remaining tables. 

The remaining four tables are separated approximately by image type. Groups of dark images are entered in the \textit{darks} table. These image groups are defined as sequential dark images with the same filter and spectral dispersion. These groups nominally contain only two images each, but can contain more in the case of, eg., testing the system or manual calibration imaging. As discussed in Section \ref{sec:current-system}, darks are only required for Broadband and K-band observations. Therefore, while the typical sequence for obtaining calibration images includes taking high spectral resolution darks in the J- and H-band, only K-band and Broadband darks are grouped and logged in the \textit{darks} table. Ignored dark images will still appear in the \textit{CHARISData} table, but will not undergo sorting in preparation for image calibration.

Similar groups of flat images are entered in the \textit{buildcal} table. Broadband and K-band flats are also paired with the nearest dark group of the same filter; this corresponding dark group is noted in the flat group's entry in the \textit{buildcal} table. Each entry in the \textit{buildcal} table corresponds to an instance of calibration images that have been run through the CHARIS DRP's \texttt{buildcal} routine. Therefore, the location of the resulting built calibration files are also denoted here in the entry for the utilized image groups.

   \begin{figure} [ht]
   \begin{center}
   \begin{tabular}{c} 
   \includegraphics[width=0.95\linewidth]{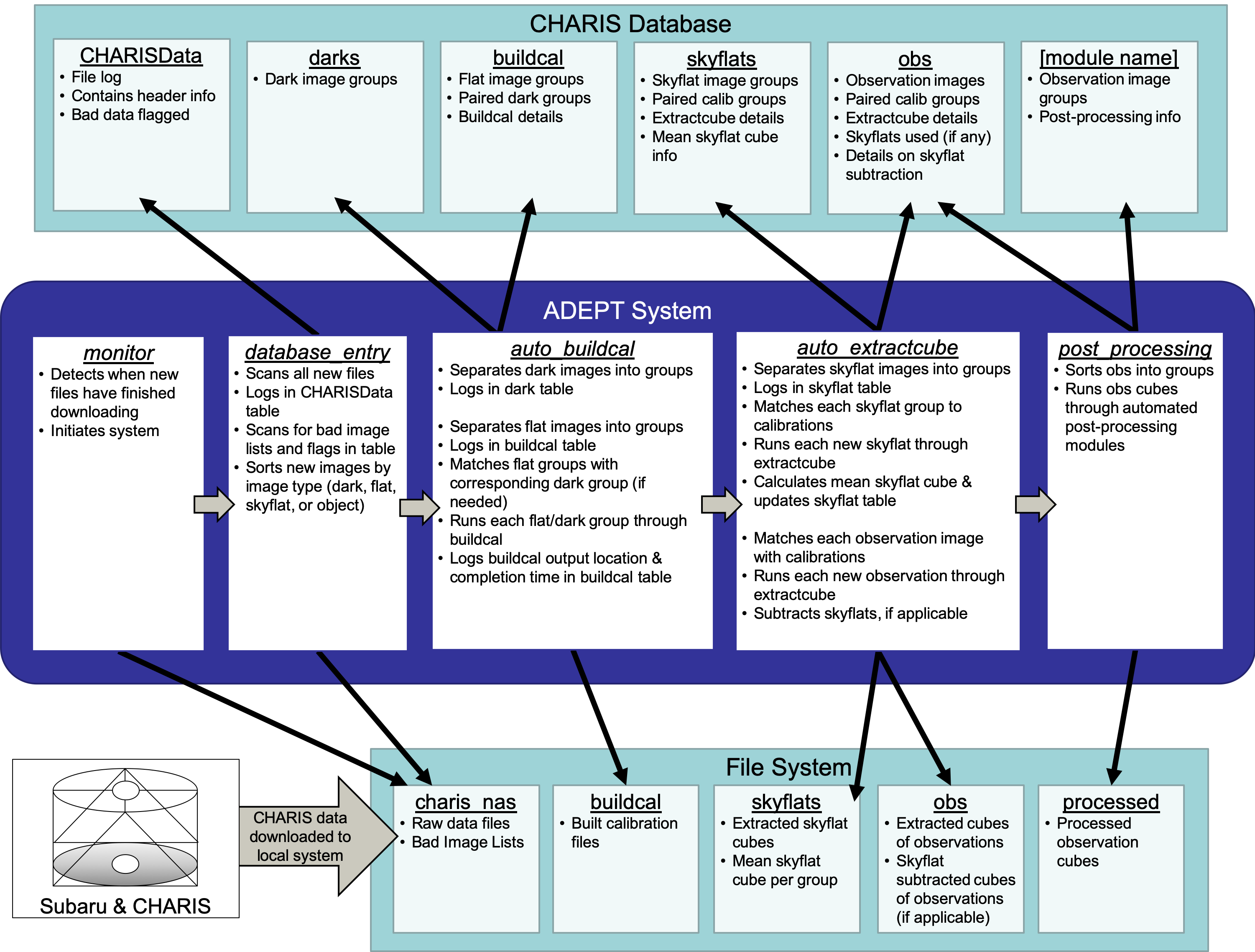}
   \end{tabular}
   \end{center}
   \caption{ \label{fig:flowchart} Diagram of the ADEPT System. The progression of the automated system is outlined in dark blue (center), where each box represents a subroutine discussed in one of the subsections of Section \ref{sec:automation}. The two sections of the new user interface that accompanies it and their interactions are shown in light blue: the organization of the file system is shown at the bottom, including raw data downloads from the telescope, with each of the major subdirectories shown (see Section \ref{ss:file-system}). The structure of the CHARIS Database is shown at the top (see Section \ref{ss:database}), with each database table indicated, along with its contents. The final ``[module name]'' table represents the tables generated by the post-processing modules, where each module creates its own database table with some custom table name (see Section \ref{ss:postprocessing}).}
   \end{figure} 
Groups of skyflat images are logged in the \textit{skyflats} table of the database. While the skyflat image groups contain more files than the flat or dark image groups, the rules that determine group membership are the same. As the data from skyflat images is processed and extracted in the same way as observations, the closest \textit{buildcal} image group of the same filter to each skyflat group is noted in the \textit{skyflats} table, as well as the location of the extracted skyflat data cubes and the resulting mean skyflat cube.  

Finally, all observation files are logged in the \textit{obs} table of the database. Unlike the other image types, observations are not yet sorted into groups at this stage. Each observation image entered in the \textit{obs} table will be listed with its corresponding calibration files, information on its data extraction, any skyflats utilized for subtraction, any post-processing used, and the location of the results.

While these five comprise the primary tables in the database, the post-processing system creates a new database table for each additional post-processing module. The names of these additional tables are specified in the module's initialization file (see Section \ref{ss:postprocessing}). As most post-processing methods require a combination of multiple exposures, observation files will be organized into groups of sequential files with the same filter in these tables. While this sorting method may not be what is in the end desired for optimal science extraction, it is sufficient for preliminary post-processing to inform future fine-tuning with the same or similar methods.

\subsection{Data Products and File System}
\label{ss:file-system}  

All of the science-grade data products produced ADEPTS are organized by file type. The ADEPTS file system is divided into five primary directories. As discussed in Section \ref{sec:current-system}, the original CHARIS NAS contains all of the raw data downloaded from CHARIS, divided by observation date.

The calibration files generated by the CHARIS DRP \texttt{buildcal} routine are stored in the \textit{buildcal} directory. Within the directory, they are organized by observation date and filter for ease of location. The exact path within the \textit{buildcal} subdirectory for a collection of flat and dark images will be noted in the group's entry in the \textit{buildcal} table of the database.

The \textit{skyflats} directory contains extracted data cubes from skyflat images, as well as the mean skyflat cube for each group. The organization within the \textit{skyflats} directory is similarly according to observation date and filter, with the exact path saved in the relevant entry of the \textit{skyflats} database table.

The \textit{obs} directory contains all extracted data cubes for observation images, as well as their skyflat-corrected data cubes, if applicable. These results are sorted within the \textit{obs} directory according to observation date, target name, and filter. The path to each image within the \textit{obs} directory is saved in the \textit{obs} table of the database. 

Finally, all output from post-processing modules is ported to the \textit{processed} directory. The results are sorted within the directory first according to the path specified for the particular post-processing method. This path is set within the initialization file for that method (see Section \ref{ss:postprocessing}). Within the subdirectory for a given routine, results are stored with the same organizational structure as used in the \textit{obs} directory, with the full subdirectory path for an observational image group denoted in the database table corresponding to the post-processing module.

\section{AUTOMATED DATA EXTRACTION, PROCESSING, AND TRACKING}
\label{sec:automation} 

The processing system at the heart of ADEPTS is designed to automate the data extraction and reduction process while fully populating the new CHARIS image database. It consists of (i) a new file monitor, (ii) image sorting and database entry, (iii) calibration and (iv) data cube extraction with the CHARIS DRP, and (v) a modular post-processing data reduction system.

The completed ADEPT System will be housed on a 72-CPU Intel\textsuperscript{\tiny\textregistered} Xenon\textsuperscript{\tiny\textregistered} Gold 6254 Processor at the University of Notre Dame. As detailed in Sections \ref{ss:autobuildcal} and \ref{ss:autoextractcube}, the automated calibration and data extraction with the CHARIS DRP is performed using multiprocessing to significantly reduce the turnaround time to science-quality images.

The number of CPUs used for the parallelization is set in the ADEPTS initialization file. This file is also used to set the full paths to the ADEPTS file system subdirectories discussed in Section \ref{ss:file-system}, paths to log files for ADEPTS feedback, database table names, parameters used by the CHARIS DRP, and initialization information for the post-processing modules (see Section \ref{ss:postprocessing}). In the event of future installation on other machines or adjustments in system infrastructure, this initialization file is the only part of the system that would need to be adjusted manually.

\subsection{The Monitor} 
\label{ss:monitor}
The ADEPTS \texttt{monitor} serves as both the first step of the automated process and its wrapper. So-called because it monitors the CHARIS NAS for any changes to the image repository using Python's \texttt{watchdog} package\footnote[1]{\url{https://pythonhosted.org/watchdog/}}, it activates once new CHARIS data downloads are complete. The monitor then gathers a list of all new image files and initiates the rest of the automated system, handing it the list of new files as input. Since the monitor serves as the wrapper for the system, it then calls each of the remaining subprocesses in turn.

\subsection{Database Entry}
\label{ss:databaseentry}

Once data download is complete, the monitor initializes the database entry routine, \texttt{database\_entry}, providing it with the list of new image files. This function first scans the headers of all the new fits files and logs the contents in the \textit{CHARISData} table of the database. 

As stated in Section \ref{ss:database}, the \textit{CHARISData} table also notes any images flagged as ``bad'' files by the observer. However, Subaru's FITS headers do not contain flags for image files manually marked as bad. Instead, bad file logs must be created by observers in the data repository, with up to one bad file log per date of observation. These bad file logs are a text file containing a list of file names that have been denoted as ``bad'' by the observer, and must be saved in the relevant date subdirectory of the CHARIS NAS's image repository.  The database entry routine scans these bad file logs and marks the appropriate files as bad in the \textit{CHARISData} table.

Finally, \texttt{database\_entry} sorts all of the scanned fits images by image type: darks, flats, skyflats, and observations, ignoring any files marked as ``bad''. The function returns four NumPy\cite{numpy} arrays, with one per image type. These arrays contain the file name, the file's subdirectory within the image repository, start and end MJD of the exposure, object name, filter, and whether the image was taken with high or low spectral dispersion. The new file array of the appropriate image type is handed then to the appropriate subsequent data extraction process by the \texttt{monitor}.

\subsection{Automated Buildcal}
\label{ss:autobuildcal}
As discussed in Section \ref{sec:current-system}, the CHARIS DRP processes dark and flat images with the subroutine \texttt{buildcal}. Once \texttt{database\_entry} logs new images in the database and sorts them by imagetype, the arrays of new dark and flat images are passed to the automated buildcal subroutine, \texttt{auto\_buildcal}. 

This subroutine first sorts the new K-band and Broadband dark images into groups and logs them in the \textit{darks} table of the database as described in Section \ref{ss:database}. It then sorts the new flat images into similar groups, but for all filters. K-band and Broadband flat groups are paired with the dark group of the same filter that is the closest in time, and the results are logged in the \textit{buildcal} table of the database. 

Notably, the system is designed to be robust to downloads of incomplete image groups. If files in the most recent download would be part of a previously-added image group, the old image group will be modified to include the new files and any subsequent image processing will be redone. The \textit{buildcal} database table contains a column noting the completion time of the most recent instance in which the DRP's \texttt{buildcal} function was run on the file group in question; this allows users to check if any modifications have been made to necessary calibration images between data retrieval and publication. However, data downloads are expected to occur daily, so any updates in calibration files due to incomplete image download would be unlikely to arise more than one day after the initial download. This mechanism serves primarily as a precaution against errors in the file download process.

Once all of the new and modified flat groups have been sorted and paired with their corresponding dark groups, if applicable, they are run through the CHARIS DRP's \texttt{buildcal}. By design, it is intended to be compatible with the most recent version of the CHARIS DRP available at the time it is run, such that it will be compatible with code updates as available on GitHub.\footnote[2]{\url{https://github.com/PrincetonUniversity/charis-dep}} However, to accommodate the use of flat and dark image groups, the current version of the CHARIS DRP script utilized by ADEPTS is one that has been modified to accept multiple flat and dark images to create a single array of calibration files. The \texttt{buildcal} script in use has also been modified to allow any output files to be ported to a specified directory and to accept all required parameters as initial inputs instead of providing individual prompts for each in the terminal after program initialization. The latter is required for the calibration to occur quietly without user intervention. The independent calls to the DRP's \texttt{buildcal} within in the \texttt{auto\_buildcal} package utilize \texttt{buildcal}'s parallel processing capability to speed processing time.

The calibration files generated by \texttt{buildcal} are then saved to the appropriate subdirectory for the flat image group in the \textit{buildcal} section of the file system, organized by date and filter. Finally, the \textit{dark} and \textit{buildcal} tables of the database are updated with the run's completion time. 

\subsection{Automated Extractcube}
\label{ss:autoextractcube}
Once the calibrations have been properly sorted, logged, and built, the \texttt{monitor} initiates \texttt{auto\_extractcube}, providing it with the arrays of new skyflat and observation files. This subroutine first sorts the new skyflat images into groups in a similar fashion to the darks and flats. Each group of skyflat images is then paired with the set of calibration files in the same filter that minimizes the elapsed time between the skyflat and flat exposures. The DRP's \texttt{extractcube} function is used to extract the data cube for each skyflat image, using the previously determined calibration files. These data cubes are then combined to create a single mean skyflat cube for the group. All information on the skyflat image group and extraction is logged in the \textit{skyflat} table of the database, including output path and the date and time at which the mean skyflat cube was last updated.

Once the mean skyflat for each group has been created, \texttt{auto\_extractcube} begins processing the observation files. Unlike the other image types, the observation files are regarded independently instead of belonging to image groups. As with the skyflats, each observation image is matched to the nearest calibration files of the same filter and the data cube is extracted via \texttt{extractcube}. The subroutine then searches for the skyflat group with the same filter that is nearest in time to each observation cube within some maximum time separation; this is nominally set at 12 hours, though it can be adjusted in the ADEPTS initialization file. If corresponding skyflats within this time limit are found, the mean skyflat cube will be subtracted from the observation image cube to create a new corrected image cube. All information on the observation data extraction and subsequent skyflat correction is stored in the obs table of the database, including completion time and location of the skyflat-subtracted image cube, if applicable. 

As with \texttt{buildcal}, this function is intended to be compatible with \texttt{extractcube} version updates. Initial updates to the DRP code first are required, namely the ability to call \texttt{extractcube} with all necessary parameters from within Python and the ability to port the extracted data cubes to a separate path.  Unlike \texttt{buildcal}, \texttt{extractcube} has no innate parallelization. Therefore, \texttt{auto\_extractcube} uses Python's \texttt{multiprocessing} package\footnote[3]{\url{https://docs.python.org/3/library/multiprocessing.html}} to perform the individual \texttt{extractcube} calls in parallel.

\subsection{Post-Processing}
\label{ss:postprocessing}
While not required for data cube extraction, the ADEPT System allows for the addition of PSF removal programs and other post-processing software, such as ADI, SDI, LOCI, and KLIP. Any post-processing code can be added to the post-processing subroutine as a modular Python code, provided that it can be run without human intervention. While many such processes require fine-tuning for optimal results, the automation of this process will be able to provide preliminary results for a given array of default reduction parameters. 

As an example, since IDL can be called within a Python script, ADEPTS is easily well-suited to execute the single \texttt{autoreduce} command that performs a quick-look reduction of CHARIS data in the currently IDL-based CHARIS Data Processing Pipeline\cite{Currie2018} (see paper 11448-330 in these proceedings for more details).   The run-time required for \texttt{autoreduce} to sky-subtract, register, spectrophotometrically calibrate, and PSF-subtract images and then blindly identify and extract raw (not-throughput corrected) spectra ranges between 10 and $\sim$ 15 minutes for a typical 30-minute CHARIS sequence (much shorter for a shallow sequence).   Additional calibration steps (e.g. forward-modeling, better tuned PSF subtraction) in the CHARIS DPP can likewise be scripted within ADEPTS.   All steps will be more seamless once the DPP is translated into Python 3.7 sometime in 2021 or shortly thereafter.  Thus, ADEPTS can lay the groundwork for near-automatic quick-look data reductions and science-grade reductions, approaching that implemented with the \textit{Gemini Planet Imager}\cite{Wang2018}.

To add a new post-processing module to the ADEPT System, one must create a stand-alone Python code to call the function and an initialization file. The initialization file must include the name of the Python script which will run the program, the path within the \textit{processed} file directory where output of the script should be written, and the table name that will be used to track observation groups that have undergone this processing in the database. It may also include any additional parameters required as input for the module's Python script. These additional parameters will appear in the module's database table to track the values used for each image group's most recent run.

The module's Python script itself should have a high-level function called \texttt{main} that performs all of the desired processing for the module; this is the function that will be called by \texttt{post\_processing}. The module's \texttt{main} function must accept the input path, a list of file names, and the output path as required parameters. Additional optional parameters may be accepted. Any keywords provided in the module's initialization file that are not required by \texttt{post\_processing} are provided to the module's \texttt{main} function as named parameters. The module itself is free to reference the database for any required information on the provided files. 

If post-processing modules are present when the \texttt{post\_processing} subroutine is called, it will first sort any files that have not undergone the requested post-processing procedure into groups of observations using the same organization rules as with other image types. These groups of images will then be processed as a unit through each independent module, with their utilized parameters and completion times logged in the respective database tables.

\section{CONCLUSIONS}
\label{sec:conclusions}

While progress on the ADEPTS source code is ongoing, the database creation and population (\texttt{database\_entry}), as well as sorting, building, and extracting the dark, flat, and skyflat files (\texttt{auto\_buildcal} and the skyflat handling within \texttt{auto\_extractcube}) are complete. The implementation of ADEPTS as a backend of the CHARIS data system will improve its ease of use with its improved user interface; its new SQLite database and sorted file system for science-grade data products will allow file information and data to be easily located. It will also likely reduce the time to science acquisition, as it provides automated data sorting, cross-matching, calibration, and extraction. Its housing on a 72-CPU Intel\textsuperscript{\tiny\textregistered} Xenon\textsuperscript{\tiny\textregistered} Gold 6254 Processor will allow for significant reductions in processing time through parallelization. As a whole, the ADEPTS package will allow CHARIS users to obtain science-quality data products in a significantly reduced time frame compared to current methods.

\acknowledgments    
 
Based in part on data collected at Subaru Telescope, which is operated by the National Astronomical Observatory of Japan. The authors wish to recognize and acknowledge the very significant cultural role and reverence that the summit of Mauna Kea has always had within the indigenous Hawaiian community. We are most fortunate to have the opportunity to conduct observations from this mountain.

\bibliography{main} 
\bibliographystyle{spiebib} 

\end{document}